\documentclass[conference]{IEEEtran}
\IEEEoverridecommandlockouts
\usepackage{cite}
\usepackage{amsmath,amssymb,amsfonts}
\usepackage{float}
\usepackage{algorithm}
\usepackage{algpseudocode}
\usepackage{setspace}
\usepackage{colortbl}
\usepackage{booktabs}
\usepackage{graphicx}
\usepackage{textcomp}
\usepackage{multirow}
\usepackage{array}
\usepackage{subcaption} 
\usepackage{xcolor}
\def\BibTeX{{\rm B\kern-.05em{\sc i\kern-.025em b}\kern-.08em
    T\kern-.1667em\lower.7ex\hbox{E}\kern-.125emX}}
\begin{document}

\title{
Hardware-Efficient Softmax and Layer Normalization with Guaranteed Normalization \\for Edge Devices\\
}
\author{
    Dawon Choi$^{1}$, Hana Kim$^{2}$, Ji-Hoon Kim$^{1,2}$ \\
    \IEEEauthorblockA{$^{1}$Dept. of Artificial Intelligence Semiconductor Engineering, Hanyang University, Seoul, Republic of Korea}
    \IEEEauthorblockA{$^{2}$Dept. of Electronic Engineering, Hanyang University, Seoul, Republic of Korea}
    Email: cdw4777@hanyang.ac.kr \\
}
\maketitle

\begin{abstract}
In Transformer models, non-GEMM (non-General Matrix Multiplication) operations—especially Softmax and Layer Normalization (LayerNorm)—often dominate hardware cost due to their nonlinear nature. To address this, previous approximation studies mainly target \textit{rank-oriented} tasks, which is acceptable for classification. However, edge Natural Language Processing (NLP) applications and edge generative AI are largely evaluated based on \textit{score-oriented} tasks, so normalization-guaranteed non-GEMM operations are essential.
We propose a hardware-efficient Softmax and LayerNorm with Guaranteed Normalization for Edge devices. Our design employs hardware-efficient approximation methods while preserving the normalization (Softmax: $\sum p = 1$, LayerNorm: $\sigma = 1$).
Our architecture is described in Verilog HDL and synthesized using the Samsung 28nm CMOS process. In accuracy evaluation, we achieve high accuracy with minimal degradation; GLUE +0.07\%, SQuAD -0.01\%, perplexity -0.09\%. 
Implementation results show that our architecture is small; $942\mu m^2$ for Softmax, $1199\mu m^2$ for LayerNorm. Compared to the state of the art, we achieve up to 11x and 14x reduction in area, respectively.
\end{abstract}

\begin{IEEEkeywords}
Transformer, Softmax, Layer Normalization, Hardware, Score-Oriented, Edge Devices, NLP
\end{IEEEkeywords}

\section{Introduction}
Transformers\cite{b1} are widely used for Natural Language Processing (NLP) tasks, as self-attention enables efficient modeling of long-range dependencies. Depending on the task, several architectural variants are used: encoder-only, decoder-only and encoder-decoder. 
BERT\cite{b2}, a representative encoder-only Transformer, is pre-trained with masked language modeling and fine-tuned with task-specific heads—typically for text classification (\textit{rank-oriented}). By contrast, decoder-only Transformers such as GPT\cite{b3} are trained with a causal language modeling objective to predict the next token, and are typically evaluated on language modeling benchmarks (\textit{score-oriented}).
From a hardware perspective, Transformer computations can be divided into General Matrix Multiplication (GEMM) and non-GEMM—such as Softmax, Layer Normalization (LayerNorm) and GeLU. Non-GEMM operations often become the main bottleneck because of its nonlinear operations, which are computationally expensive and challenging to implement.

\begin{figure}[!t]
\centering
\includegraphics[width=1.0\columnwidth]{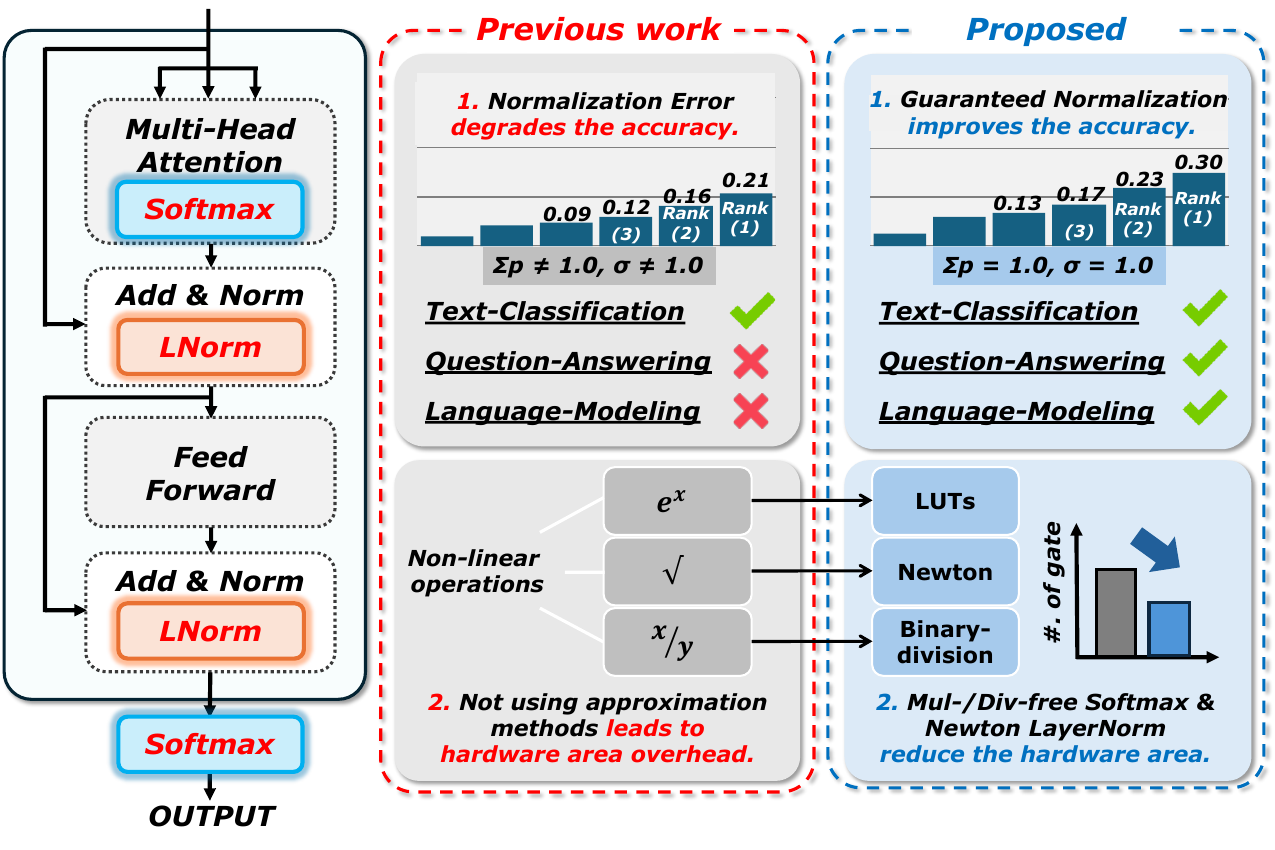}
\caption{Impact of Normalization Error in non-GEMM Operations on Score-oriented Tasks.}
\label{fig}
\vspace{-0.4cm}
\end{figure}

Accordingly, numerous studies have targeted non-GEMM to reduce hardware cost. Sole\cite{b4} reports low area overhead by adopting log-domain Softmax and a LUT-based LayerNorm, achieving up to 3.32x hardware efficiency against the baseline. Also, Softermax\cite{b5} and pseudo-Softmax\cite{b6} simplify the exponential by using the base-2 Softmax, reducing the hardware area to 0.25x. RMSNorm\cite{b7}, which eliminates the mean subtraction, is expected to reduce. \\
While these approximation methods reduce hardware area, they are predominantly \textit{rank-oriented} preserving relative ordering or top-$\mathrm{k}$ rankings, rather than maintaining the accuracy of absolute scores. Fig. 1 illustrates that this deviation in score—hereafter called \textit{normalization error}—is usually tolerable for \textit{rank-oriented} tasks as long as the ordering is preserved; however, it degrades accuracy on \textit{score-oriented} tasks. \\
To overcome these limitations, we propose \textbf{a hardware-efficient Softmax and LayerNorm with Guaranteed Normalization} for edge devices. \\
Our main contributions are:
\begin{itemize}
    \item Normalization-guaranteed non-GEMM blocks: Softmax ($\sum p = 1$) and LayerNorm ($\sigma = 1$).
    \item Multiplier-/Divider-free, hardware-friendly Softmax with look-up table-based exponential and fixed-point divider.
    \item Hardware-friendly LayerNorm via a compressed Newton reciprocal square root unit.
    \item Our design achieves both accuracy and guaranteed normalization in \textit{score-critical} tasks while minimizing hardware area overhead.
\end{itemize}
Experimental results show our design achieves high accuracy—GLUE +0.07\% and only 0.01\%/0.09\% drops on SQuAD/perplexity—while implementing Softmax and LayerNorm with a total area of $2141\mu m^2$.

\section{Background}
In this section, we review the key concepts used in this work and motivation of our design.
\subsection{Normalization error}
Let $p$ denote the Softmax output and $\sigma$ denote the standard deviation in normalized output. We define the \textit{normalization error} by $\bigl|1-\sum p\bigr|$ for Softmax and $\bigl|1-\sigma\bigr|$ for LayerNorm.

GLUE\cite{b8} is formulated as sentence- or pair-level classification, where relative logit ordering is sufficient. Therefore, normalization errors in Softmax and LayerNorm have limited impact on GLUE accuracy.
By contrast, \textit{score-oriented} tasks such as SQuAD\cite{b9} and perplexity\cite{b10} are sensitive to normalization. In the BERT QA head\cite{b2}, start and end logits are converted to probability distributions via Softmax and spans are scored using these log-probabilities. 
Similarly, Eq. (1) defines the perplexity\cite{b10} for test set $W = w_1w_2 . . . w_N$ as the inverse probability of the test set, normalized by the number of words.
Therefore, normalization error becomes critical in these \textit{score-oriented} tasks.
\begin{equation}
    \label{eq:placeholder_label}
    \text{perplexity}(W) = P(w_1 w_2 \ldots w_N)^{-\frac{1}{N}} = \sqrt[N]{\frac{1}{P(w_1 w_2 \ldots w_N)}}
\end{equation}

\subsection{Nonlinear operations}
\begin{equation}
    \text{Softmax}(X_i) = \frac{\exp(X_i - X_{\max})}{\sum_{j=1}^{n} \exp(X_j - X_{\max})}
\end{equation}
\begin{equation}
    \begin{array}{c}
\text {LayerNorm}\left(X_{i}\right)=\frac{X_{i}-\mu}{\sigma} \times \gamma+\beta \\
\text {where } \mu=\frac{1}{N} \sum_{i=0}^{N} X_{i} \text { and } \sigma=\sqrt{\frac{1}{N} \sum_{i=0}^{N}\left(X_{i}-\mu\right)^{2}}
\end{array}
\end{equation}

In non-GEMM, the exponential (Eq. (2)), division and square root (Eq. (3)) dominate the area overhead. 
Although existing approximations reduce hardware cost, they often increase the normalization error.

\subsection{Motivation}
Fig. 2 illustrates the relationship between the normalization error and approximation level. In Softmax, probability normalization enforces $\sum p = 1$ and unit-variance normalization targets $\sigma = 1$ for LayerNorm. Increasing the approximation level typically increases normalization error, while decreasing it reduces error but incurs higher area overhead. Therefore, maintaining an optimal balance between such normalization and approximation is crucial for non-GEMM efficiency.

\section{Proposed Design and Implementation}
We propose hardware-efficient Softmax and LayerNorm with guaranteed normalization. This section presents our algorithms and hardware architectures.
\begin{figure}[!t]
\centering
\includegraphics[width=1.0\columnwidth]{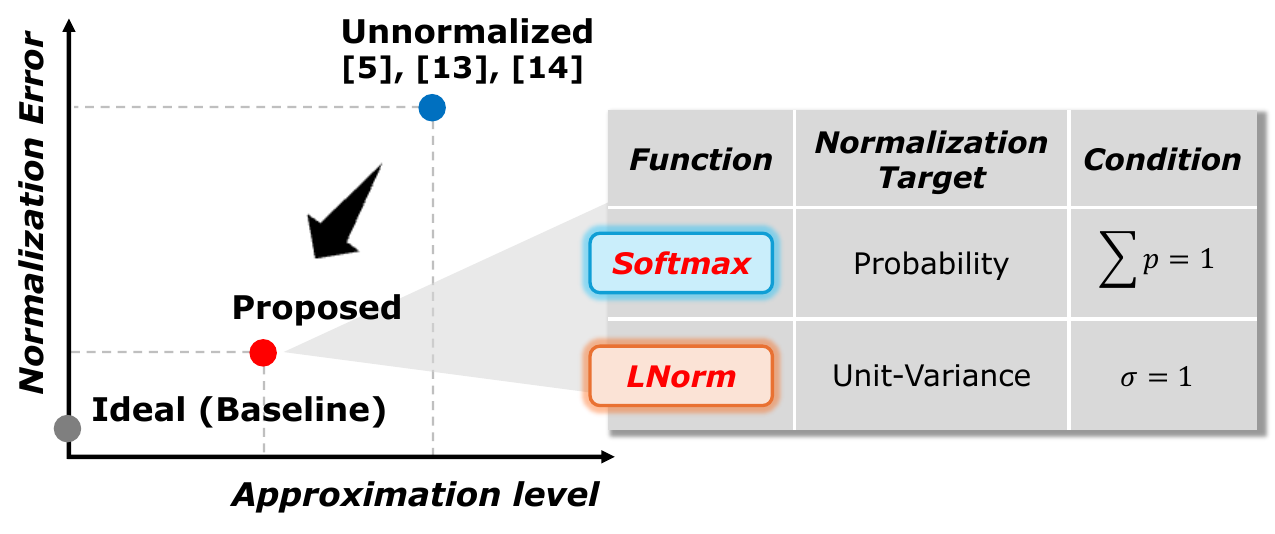}
\caption{Normalization Error ($\bigl|1-\sum p\bigr|,\ \bigl|1-\sigma\bigr|$) vs. Approximation Level.}
\label{fig}
\vspace{-0.3cm}
\end{figure}

\begin{algorithm}[!t]
\caption{Proposed Mul-/Div-free Softmax}
\scalebox{0.9}{
\parbox{1.1\columnwidth}{
\begin{algorithmic}[1]
\State \textbf{Input:} $X \in \mathbb{R}^N,$ \textbf{Output:} $Y \in \mathbb{R}^N$
\Statex
\State $\Delta_i \gets \max(X) - X_i$ \Comment{Finding maximum value}
\State $\mathrm{frac}_i \gets \left\lfloor \Delta_i / R \right\rfloor$ 
\State $\mathrm{rem}_i \gets \Delta_i - R \cdot \mathrm{frac}_i$ 
\Statex
\State $a_i \gets \exp(-\mathrm{frac}_i)$ \Comment{Coarse exponential}
\State $b_i \gets \exp(-\mathrm{rem}_i)$ \Comment{Residual exponential}
\State $y_i \gets a_i \cdot b_i$ \Comment{LUT-based exponential approximation}
\Statex
\For{$i \gets 1$ \textbf{to} $N$}
    \State $Z \gets Z + y_i$
\EndFor
\State $Y \gets \text{FxP\_Div}(y,\ Z)$ \Comment{Division approximation}
\State \textbf{return} $Y$
\end{algorithmic}
}
}
\end{algorithm}

\begin{algorithm}[!t]
\caption{Proposed Newton LayerNorm}
\scalebox{0.9}{
\parbox{1.1\columnwidth}{
\begin{algorithmic}[1]
\State \textbf{Input:} $X \in \mathbb{R}^N,$ \textbf{Output:} $Y \in \mathbb{R}^N$ 
\Statex
\For{$i \gets 0$ \textbf{to} $C$}
    \State $E_x \gets E_x + X_i$        
    \State $E_{x^2} \gets E_{x^2} + X_i * X_i$    
\EndFor
\State $E_x \gets E_x / C, \ E_{x^2} \gets E_{x^2} / C $     \Comment{Mean calculation}
\State $var \gets E_{x^2} - {E_x}^2 $     \Comment{Variance calculation}
\Statex
\State $\mathrm{mean} \gets E_x$
\State $\mathrm{std} \gets \textit{CoRN-LN}(var)$ \Comment{Reciprocal Newton method}
\Statex
\State $Y \gets (X - mean) / std$
\State \textbf{return} $Y$
\end{algorithmic}
}
}
\end{algorithm}

\begin{table*}[!t]
\caption{Results in Natural Language Processing Tasks.}
\label{tab:nlp-results}
\resizebox{\textwidth}{!}{
\begin{tabular}{lcccccccccccc}
\toprule
\midrule
& \multicolumn{9}{c}{\textbf{Natural Language Processing in GLUE}} & \multicolumn{2}{c}{\textbf{SQuAD v1.1}} & \multicolumn{1}{c}{\textbf{perplexity}} \\
\cmidrule(lr){2-10}\cmidrule(lr){11-12}\cmidrule(lr){13-13}
\textbf{TASK} & MRPC & SST-2 & CoLA & QNLI & MNLI & QQP & RTE & STS-B & WNLI & Eval\_F1 & Eval\_Exact & perplexity \\
\midrule
FP32        & 90.24 & 93.23 & 58.80 & 91.25 & 84.69 & 87.88 & 68.95 & 88.43 & 56.34 & 90.50 & 83.46 & 13.72 \\
\textbf{FP32 + Ours} & 90.24 & 93.23 & 59.05 & 91.29 & 84.71 & 87.91 & 68.95 & 88.45 & 56.34 & 90.49 & 83.49 & 13.74 \\
\midrule
\bottomrule
\end{tabular}}
\vspace{-0.35cm}
\end{table*}

\begin{table}[!t]
\renewcommand{\arraystretch}{0.9}
\fontsize{9.5pt}{11.5pt}\selectfont
\centering
\caption{Accuracy Comparison for Score-Oriented Tasks}
\label{tab:nlp-results}
\begin{tabular}{l cccc}
\toprule
\midrule
\textbf{TASK} & \cite{b5}* & \cite{b13}** & \cite{b14} &  \text{\textbf{Proposed}}\\
\midrule
\textbf{SQuAD (\%)}      & -0.49 & -0.68 & -- & -0.01 \\
\textbf{perplexity (\%)} & -- & -13.68 & -0.73 & -0.09 \\
\midrule
\bottomrule
\end{tabular}
\vspace{-0.1cm}
\begin{flushleft}
\footnotesize
\textit{\ \ * includes only Softmax} \\
\textit{\ \ ** includes Softmax, LayerNorm and GeLU}
\end{flushleft}
\vspace{-0.6cm}
\end{table}

\subsection{Algorithm: Softmax}
Algorithm 1 presents the proposed multiplier (mul)-/divider (div)-free Softmax algorithm, which uses approximation methods while guaranteeing probability normalization.\\
\textbf{Approximation method (Exponential-LUTs):}
Inputs are first stabilized by subtracting the maximum, so $e^{-q}$ lies in $(0,\ 1]$. Due to the exponential decay, most values become negligible after stabilization. By decomposing $q=R\cdot frac + rem$, the exponential can be factorized into a coarse term and a residual term, where the residual follows a geometric ratio and has a fixed domain independent of the coarse component. As shown in Eq. (4), this yields a two-term factorization of $e^{-q}$.
\begin{equation}
    e^{-q} = e^{-R\cdot frac} \cdot e^{-rem}
\end{equation}
\textbf{Approximation method (Fixed-Point Division):}
The fixed-point divider used in hardware is described in Section III-C.

\subsection{Algorithm: Layer Normalization}
Algorithm 2 presents the proposed Newton LayerNorm algorithm that approximates the square root based on Newton method\cite{b11}, ensuring the unit-variance normalization.\\
\textbf{Approximation method (Reciprocal Newton):}
We compute an approximation based on Newton method with a Leading-One-Detector (LOD)-based initial guess and reformulated reciprocal form\cite{b12}, as represented in Eq. (5).
\begin{equation}
    x_{i+1}={0.5\,\big\lfloor x_i + 1/(x_i\cdot n) \big\rfloor}
\end{equation}

\subsection{Architecture: Softmax}
As shown in Fig. 3, we propose a Mul-/Div-free Softmax: exponential is factorized with two look-up table (LUT)s, and normalization uses a custom fixed-point divider (FxP\_Div). The datapath has 3 stages: (i) max-subtraction, (ii) exponential, and (iii) normalization.\\
\textbf{Hardware Implementation (Two Exponential-LUTs):}
As discussed in Section III-A, the coarse and residual terms are computed using two LUTs indexed by $frac$ and $rem$, respectively. We select a radix of $R = 8$ to accurately approximate exponential functions in the lower input range, enabling two compact LUTs with 7 and 8 entries, respectively.\\
\textbf{Hardware Implementation (Fixed-Point Division):}
FxP\_Div is a fixed-point divider based on a binary shift–subtract algorithm, implemented using hardware-friendly blocks. Instead of performing a general division, FxP\_Div exploits the normalization condition to approximate the reciprocal scaling factor \(D_{\max}/Z\), where \(Z=\sum y\) and \(D_{\max}=2^{bit}\). So, the resulting fixed-point scaling factor is used to rescale each \(y\), producing normalized outputs through shift-and-add operations.

\begin{figure}[!t]
\centering
\includegraphics[width=1.0\columnwidth]{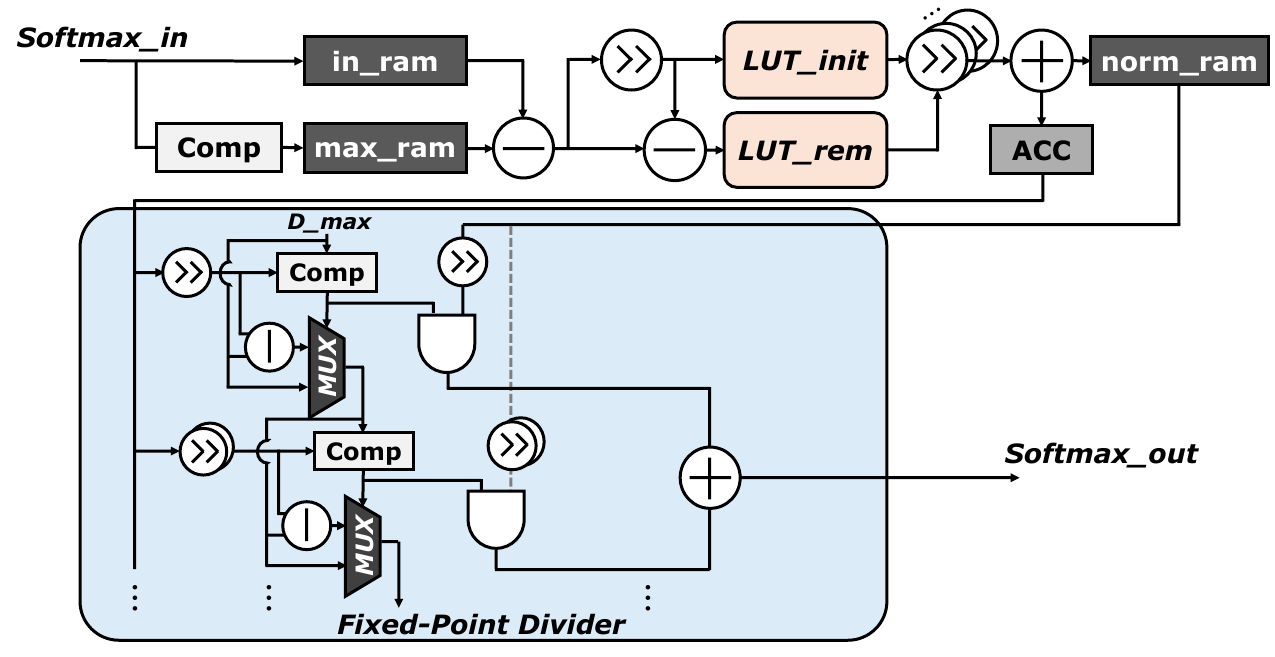}
\caption{Proposed Softmax Architecture.}
\label{fig}
\vspace{-0.4cm}
\end{figure}

\begin{figure}[!t]
\centering
\includegraphics[width=1.0\columnwidth]{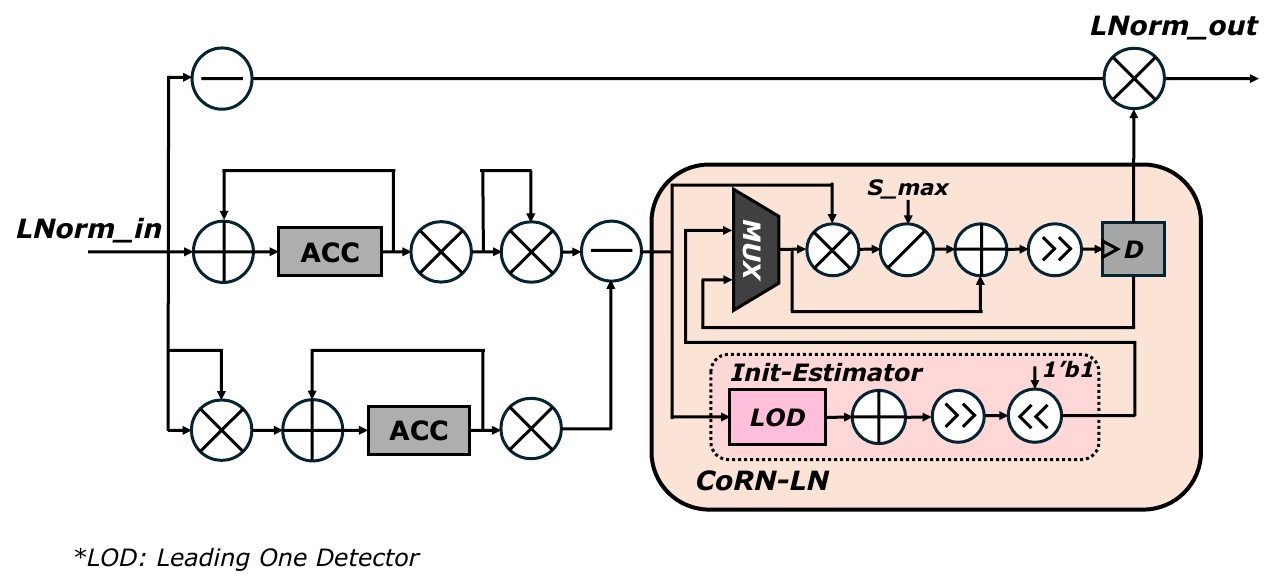}
\caption{Proposed LayerNorm Architecture.}
\label{fig}
\vspace{-0.45cm}
\end{figure}

\subsection{Architecture: Layer Normalization}
As shown in Fig. 4, proposed LayerNorm operates in 2 stages; (i) mean and variance computation, and (ii) normalization. We replace the costly square root with a compact approximation method. \\
\textbf{Hardware Implementation (Reciprocal Newton):}
Ref.\cite{b12} adopts Newton method for LayerNorm, which provides higher accuracy. In addition, the LOD-aware initial guess enables the estimation to complete in only 2-cycle and by reformulating the formula in reciprocal square root form, the output stage uses a multiplier instead of a divider.

\begin{table*}[!t]
\centering
\caption{Softmax and LayerNorm Architecture Comparison}
\label{tab:comparison_top}
\footnotesize
\renewcommand{\arraystretch}{1.1}
\begin{tabular}{|l|c|c|c|c|c|}
\hline
& \textbf{  SCIS'24 \cite{b15}  } & \textbf{TCAS-II'25 \cite{b16}} & \textbf{TCAS-I'25 \cite{b17}} & \textbf{TCAS-II'20 \cite{b18}} & \textbf{  \textit{Proposed}  } \\
\hline
\textbf{Precision} & INT19 & INT32 & INT8 & Fixed-Point16 & INT8 \\
\hline
\textbf{Technology node} & 65nm & 28nm & 90nm & 28nm & 28nm \\
\hline
\textbf{Max Frequency (GHz)} & 0.2 & 1.0 & 1.0 & 0.5 & 1.0 \\
\hline
\multirow{1}{*}{\textbf{Area ($\mu m^2$)}} & & & & & \\
\quad \textit{Softmax} & 2492 & -- & 4234 & 10081 & 942 \\
\quad \textit{LayerNorm} & 17388 & 3684 & -- & -- & 1199 \\
\hline
\multirow{1}{*}{\textbf{Latency (cycle, \textit{N} inputs)}} & & & & & \\
\quad \textit{Softmax} & \textit{N} & -- & \textit{N} & \textit{N} & \textit{N} \\
\quad \textit{LayerNorm} & \textit{N} & \textit{N} & -- & -- & \textit{N+1} \\
\hline
\textbf{Normalization Guarantee (Status)} &
{ Unnormalize } &
{ Unnormalize } &
\textbf{ Guarantee } &
\textbf{ Guarantee } &
\textbf{ Guarantee } \\

\hline
\end{tabular}
\vspace{-0.4cm}
\end{table*}

\section{Experimental Results}
We evaluate the impact of normalization on accuracy and hardware cost through software and hardware experiments in this section.
\subsection{Experimental Setup}
In Section IV-B, we assess Softmax and LayerNorm on NLP tasks—GLUE\cite{b8}, SQuAD v1.1\cite{b9}, and perplexity\cite{b10}—described in PyTorch using FP32. We additionally evaluate GPT-style perplexity to analyze the accumulation of normalization errors in autoregressive decoding.
In Section IV-C, the proposed architecture is described in Verilog HDL and synthesized using the Samsung 28nm CMOS process, targeting an operating frequency of $250MHz$.

\subsection{Accuracy Evaluation}
We evaluate our method on GLUE and SQuAD v1.1 using BERT-base, and perplexity on WikiText using GPT-Neo 1.3B. As shown in Table I, results of FP32+Ours match or improve the FP32 (baseline, ideal) across all GLUE tasks. Importantly, no task degrades with an average improvement of 0.07\%.\\
On SQuAD v1.1, our model achieves 90.49 F1 and 83.49 Exact Match, showing effectively unchanged, from the baseline (\textbf{-}0.01\% F1 / +0.04\% EM) due to guaranteed normalization. In contrast, unnormalized non-GEMM methods report drops of 0.49\% and 0.68\% in Refs.~\cite{b5,b13}.
Perplexity, highly sensitive to normalization error, shows minimal degradation: 13.72 (baseline) vs. 13.74 (ours), only 0.09\% degradation. In contrast, Ref.\cite{b13} observes a 13.68\% drop and 3.29\% drop when approximating only Softmax and GeLU—implying LayerNorm is the dominant bottleneck of accuracy degradation. In contrast, our normalization-guaranteed Non-GEMM design preserves accuracy. In addition, Ref.\cite{b14} explores precision sweeps; even at their optimal point, perplexity drops by 0.73\%. 
Furthermore, Fig. 5 illustrates the distribution of normalization error for Softmax and LayerNorm measured in perplexity evaluation. Our proposed operations show error nearly identical to the baseline (ideal) across the entire distribution. Notably, 77.1\% of Softmax and 100\% of LayerNorm error values fall within near-zero range below $0.2\times 10^{-6}$.
These results demonstrate that our normalization-guaranteed non-GEMM approximations can preserve end-to-end NLP quality.

\subsection{Implementation Results}
We compared our proposed architecture against four references in Table III, implemented using different technology nodes and platforms.
Ref.~\cite{b15} replaces division and square root using a small LUT and shifter, reporting $2492,\mu m^2$ for Softmax and $17388,\mu m^2$ for LayerNorm with $N$-cycle latency.
Ref.\cite{b16} uses a dynamic quantization algorithm based on integer arithmetic for LayerNorm, which leads to increased normalization error, resulting in $3684\mu m^2$ with $N$ cycles.
\begin{figure}[!t]
\centering
\includegraphics[width=1.0\columnwidth]{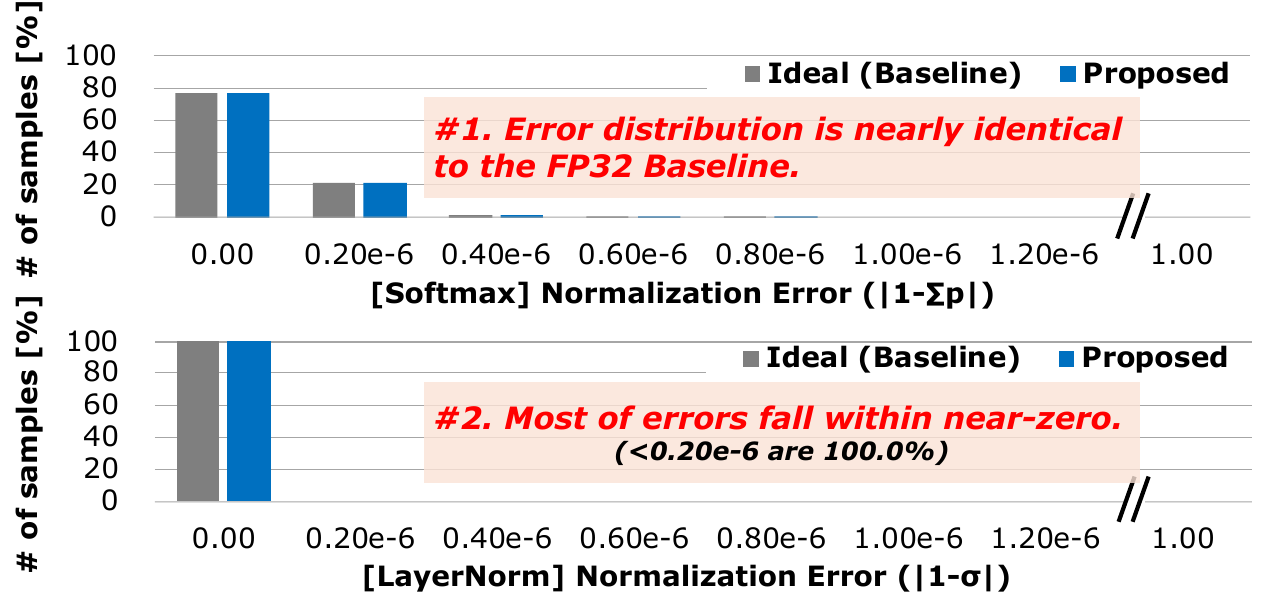}
\caption{Distribution of Softmax and LayerNorm Normalization Error in Perplexity Evaluation.}
\label{fig}
\vspace{-0.55cm}
\end{figure}
Ref.\cite{b17} approximates the exponential using $2^x$ and $4^x$, applying the same scheme to numerator and denominator via shifters and small LUTs, obtaining $4234\mu m^2$ and $N$ cycles.
Ref.\cite{b18} implements Softmax using a base-2 approach with various precision configurations, observing $10081\mu m^2$ and $N$ cycles.

In contrast, our method achieves lower hardware area; $942\mu m^2$ and $1199\mu m^2$ for Softmax and LayerNorm up to 11x and 14x smaller compared with references. For Softmax, two exponential-LUTs and a fixed-point divider minimize the hardware cost. Also, for LayerNorm, we replace the conventional square root operations with a low-cost module based on Newton method. The latency is $N+1$ cycles—an acceptable overhead for guaranteed normalization.

\section{Conclusion}
We propose hardware-efficient Softmax and LayerNorm with guaranteed normalization. Experiments on GLUE, SQuAD v1.1, and perplexity show that our design achieves high accuracy by reducing normalization error. Our design occupies $942\mu m^2$ for Softmax and $1199\mu m^2$ for LayerNorm, achieving area reductions up to 11x and 14x compared with references.
These properties make our work particularly suitable for edge NLP applications and on-device generative AI, where tight hardware budgets and high accuracy are required.

\section*{Acknowledgments}
This work was partly supported by the R\&D Program of the Ministry of Trade, Industry, and Energy (MOTIE) and Korea Evaluation Institute of Industrial Technology (KEIT) (RS-2023-00232192) and  partly supported by Institute of Information \& communications Technology Planning \& Evaluation (IITP) under the artificial intelligence semiconductor support program to nurture the best talents (IITP-(2025)-RS-2023-00253914) grant funded by the Korea government (MSIT). The EDA tool was supported by the IC Design Education Center (IDEC), Republic of Korea.


\begin{thebibliography}{00}
\bibitem{b1} K. Han, \textit{et al.}, “Transformer in transformer," in Advances in Neural Information Processing Systems (NeurIPS), vol. 34, pp. 15908–15919, 2021.
\bibitem{b2} M. V. Koroteev, “BERT: a review of applications in natural language processing and understanding,” arXiv preprint, arXiv:2103.11943, 2021.
\bibitem{b3} A. Radford, \textit{et al.}, “Improving language understanding by generative pre-training,” 2018.
\bibitem{b4} W. Wang, \textit{et al.}, “Sole: Hardware-software co-design of softmax and layernorm for efficient transformer inference,” in IEEE/ACM Int. Conf. on Computer Aided Design (ICCAD), 2023.
\bibitem{b5} J. R. Stevens, \textit{et al.}, “Softermax: Hardware/software co-design of an efficient softmax for transformers,” in IEEE/ACM Design Automation Conf. (DAC), pp. 469–474, 2021.
\bibitem{b6} G. C. Cardarilli, \textit{et al.}, “A pseudo-softmax function for hardware-based high speed image classification,” Scientific Reports, vol. 11, no. 1, pp. 15307, Jul. 2021.
\bibitem{b7} B. Zhang, \textit{et al.}, “Root mean square layer normalization,” in \textit{Advances in Neural Information Processing Systems}, vol. 32, 2019.
\bibitem{b8} A. Wang, \textit{et al.}, “GLUE: A multi-task benchmark and analysis platform for natural language understanding,” arXiv preprint, arXiv:1804.07461, 2018.
\bibitem{b9} P. Rajpurkar, \textit{et al.}, “Squad: 100,000+ questions for machine comprehension of text,” arXiv preprint, arXiv:1606.05250, 2016.
\bibitem{b10} T. Brown, \textit{et al.}, “Language models are few-shot learners,” in Advances in Neural Information Processing Systems (NeurIPS), vol. 33, pp. 1877–1901, 2020.
\bibitem{b11} R. F. King, \textit{et al.}, “The logarithmic error and Newton's method for the square root,” Communications of the ACM , vol. 12, no. 2, pp. 87–88, Feb. 1969.
\bibitem{b12} D. Choi, \textit{et al.}, “CoRN-LN: Compressed Reciprocal Newton Method for Efficient Layer Normalization," 2025 IEEE Asia Pacific Conference on Circuits and Systems (APCCAS), 2025.
\bibitem{b13} W. Wang, \textit{et al.}, “Improving Transformer Inference Through Optimized Non-Linear Operations With Quantization-Approximation-Based Strategy,” IEEE Transactions on Computer-Aided Design of Integrated Circuits and Systems, 2024.
\bibitem{b14} M. Rakka, \textit{et al.}, “SoftmAP: Software-Hardware Co-design for Integer-Only Softmax on Associative Processors,” in Design, Automation \& Test in Europe Conference (DATE), 2025.
\bibitem{b15} W. Li, \textit{et al.}, “Hardware-oriented algorithms for softmax and layer normalization of large language models,” \textit{Science China Information Sciences}, vol. 67, no. 10, pp. 200404, Oct. 2024. 
\bibitem{b16} H. Shao, \textit{et al.}, “An Efficient Layer Normalization Training Module With Dynamic Quantization for Transformers,” IEEE Transactions on Circuits and Systems II: Express Briefs, Sep. 2025.
\bibitem{b17} Y. Wu, \textit{et al.}, “MBS: A High-Precision Approximation Method for Softmax and Efficient Hardware Implementation,” IEEE Transactions on Circuits and Systems I: Regular Papers, Jul. 2025.
\bibitem{b18} D. Zhu, \textit{et al.}, “Efficient precision-adjustable architecture for softmax function in deep learning,” IEEE Transactions on Circuits and Systems II: Express Briefs, vol. 67, no. 12, pp. 3382–3385, Dec. 2020.

\end{thebibliography}
\end{document}